\documentclass[12pt]{article}
\usepackage{graphicx}
\usepackage{amsmath}
\usepackage{amssymb}
\usepackage{a4wide}

\begin{document}

\markboth{Engel and Trebin}{Stability of the decagonal quasicrystal in the
  Lennard-Jones-Gauss system}

\title{Stability of the decagonal quasicrystal in the Lennard-Jones-Gauss
  system}

\author{Michael Engel\footnote{Author for correspondence. Email:
  mengel@itap.uni-stuttgart.de}~ and Hans-Rainer Trebin\\
  {\normalsize Institut f\"ur Theoretische und Angewandte Physik,}
  {\normalsize Universit\"at Stuttgart,} \\
  {\normalsize Pfaffenwaldring 57, D-70550 Stuttgart, Germany}\\
}

%
%
%

\maketitle

\begin{abstract}

  Although quasicrystals have been studied for 25 years, there are many open
  questions concerning their stability: What is the role of phason
  fluctuations? Do quasicrystals transform into periodic crystals at low
  temperature? If yes, by what mechanisms? We address these questions here for
  a simple two-dimensional model system, a monatomic decagonal quasicrystal,
  which is stabilized by the Lennard-Jones-Gauss potential in thermodynamic
  equilibrium. It is known to transform to the approximant Xi, when cooled
  below a critical temperature. We show that the decagonal phase is an
  entropically stabilized random tiling. By determining the average particle
  energy for a series of approximants, it is found that the approximant Xi is
  the one with lowest potential energy.

\end{abstract}

\section{Introduction}

Quasicrystals are long-range ordered materials without translational symmetry.
Their structure is best understood as a decorated tiling. Each vertex of the
tiling is an integer multiple of a finite set of basis vectors. As a
consequence, the tiling can be projected from a periodic lattice in a
higher-dimensional configuration space. The subset of the lattice used for the
projection is called the de Bruijn surface.

Two scenarios for the stability of quasicrystals are currently under
discussion~\cite{Henley2006}: the perfect tiling and the random tiling.  In
both of them, the atoms are restricted to a discrete set of equilibrium
positions given by the underlying tiling structure. In the perfect tiling the
de Bruijn surface is straight and rigid. As a consequence, the atom positions
are strongly ordered. In contrast, the random tiling has an enhanced
flexibility allowing phason fluctuations of the de Bruijn
surface~\cite{Henley1991}.  Both scenarios lead to similar diffraction
patterns with identical Bragg peak positions. However in the case of the
random tiling, additional diffuse scattering is present in the background.

The deviation of a general de Bruijn surface from the straight one is
described by the phason displacement $w(r)$~\cite{Levine1985}. A constant
phason displacement does not cost energy. Thus, the free energy can be written
as a function of the phason strains $\chi_{ij}=\partial w_{i}/\partial r_{j}$.
Because elementary tiling defects of the perfect tiling are matching rule
violations whose number is proportional to the phason strain, the free energy
is linear: $F(\chi)\varpropto|\chi|$. This shows that the perfect tiling is
stabilized energetically. On the other hand, the stabilization of the random
tiling is caused by the entropy of the phason fluctuations.  The free energy
is then analytic, in lowest order $F(\chi)\varpropto\chi^{2}$.

Since the importance of entropy increases with temperature, it is assumed that
the random tiling is the preferred state at high $T$.  Upon lowering the
temperature, a locking transition to the perfect tiling has been
suggested~\cite{Ishii1989}.  Interestingly, this transition cannot appear in
two dimensions, because thermal roughening of the de Bruijn surface occurs at
all $T>0$~\cite{Tang1990}.  Furthermore, a phase transition to a periodic
crystal as energetic ground state is possible. If the structure of the crystal
is similar to the quasicrystal, in which case it is called an approximant,
then the transformation has to involve a rotation of the de Bruijn surface.
However, a full transformation via such a rotation is inefficient for large
quasicrystals, because the required total number of flips increases faster
with the size of the system than the number of atoms.

A more efficient transformation mechanism is the formation of a twinned
nanodomain structure~\cite{Steurer2005}.  The twinning corresponds to a
folding of the de Bruijn surface into piecewise straight segments, while
leaving invariant the average orientation of the surface. Note that each
approximant can appear in different orientations according to the symmetry of
the quasicrystal. Indeed, experimental observations reveal that the ground
state for many systems is a twinned approximant~\cite{Steurer2005}. However it
is not known, whether quasicrystals can be stable down to 0~K. If the phase
transition occurs at low-temperatures, the flips will be frozen in.

Currently, there are still no realistic models for studying the transition to
and from the quasicrystal state in simulations. The reason is the structural
complexity and the necessity to simulate over long times in order to reach
thermal equilibrium. Hence, we resort to a simple two-dimensional system of
identical particles interacting with the Lennard-Jones-Gauss (LJG) potential.
As far as we know, the LJG system is the only system that allows to observe a
transition between quasicrystal and approximant in simulations.

\section{Model system and computational methods}

Motivated by the existence of Friedel oscillations in the effective pair
potentials of many metals, we have recently introduced the LJG interaction
potential~\cite{Engel2007} of the form
\begin{equation}\label{equation1}
  V(r) = \frac{1}{r^{12}} - \frac{2}{r^{6}} -
  \epsilon\exp\left(-\frac{(r-r_{0})^{2}}{2\sigma^{2}}\right).
\end{equation}
The potential consists of a Lennard-Jones term and an additional Gaussian
minimum at position $r_{0}$, with depth $\epsilon$, and width given by
$\sigma$. Its phase diagram in the $r_{0}$-$\epsilon$ plane is surprisingly
rich: several crystals and two quasicrystals can be grown from the melt in
numerical simulations.

In the following, the parameters $r_{0}=1.52$, $\epsilon=1.8$, and
$\sigma^{2}=0.02$ are fixed.  This stabilizes a decagonal quasicrystal at
elevated temperatures.  As shown in \cite{Engel2007}, a reversible phase
transition to the approximant Xi occurs upon lowering the temperature below
$T_{c}=0.37\pm0.03$ ($k_{B}=1$). We note that the latter is not the $T=0$
ground state. There is a hexagonal phase, which is still more stable at low
temperatures. However, we are not concerned with this phase, since its
nucleation radius is too big. Although for smaller $r_{0}$ values the
approximant Xi can be made stable down to $T=0$, we chose $r_{0}=1.52$,
because the formation of the decagonal phase seems to be easiest and fastest
there.

Simulations are run with a combination of molecular dynamics and Monte Carlo.
Molecular dynamics is used to relax phonon strains. Periodic boundary
conditions and a Nose-Hoover thermo-/barostat for constant temperature and
pressure $P=0$ are applied. The potential is cut off at $r=2.5$. Monte Carlo
helps to speed up the phason strain relaxation. This is necessary, because the
lowest energy approximant sometimes do not have a strictly straight, but
slightly modulated de Bruijn surface. A single MC step consists of a random
displacement of particles inside a ring with radius $r\in [0.50,0.66]$, which
is the flip distance of the quasicrystal.

\section{The decagonal random tiling}

Above $T_{c}$, the decagonal random tiling is the thermodynamic equilibrium
state. At $T=0.50$, which is $90\%$ of the melting temperature, in average
$3\cdot10^{-5}$ flips per particle and time step are observed in molecular
dynamics simulations. The number has to be compared with the average vibration
time of a particle in its local potential minimum, which is of the order of
$10-100$ MD steps. This means that there is about one phason flip per $1000$
phonon vibrations. The flips are not correlated in space/time, although the
positions of possible flips are of course restricted by the tiling structure.

\begin{figure}[t]
  \centering
  \includegraphics[width=0.95\textwidth]{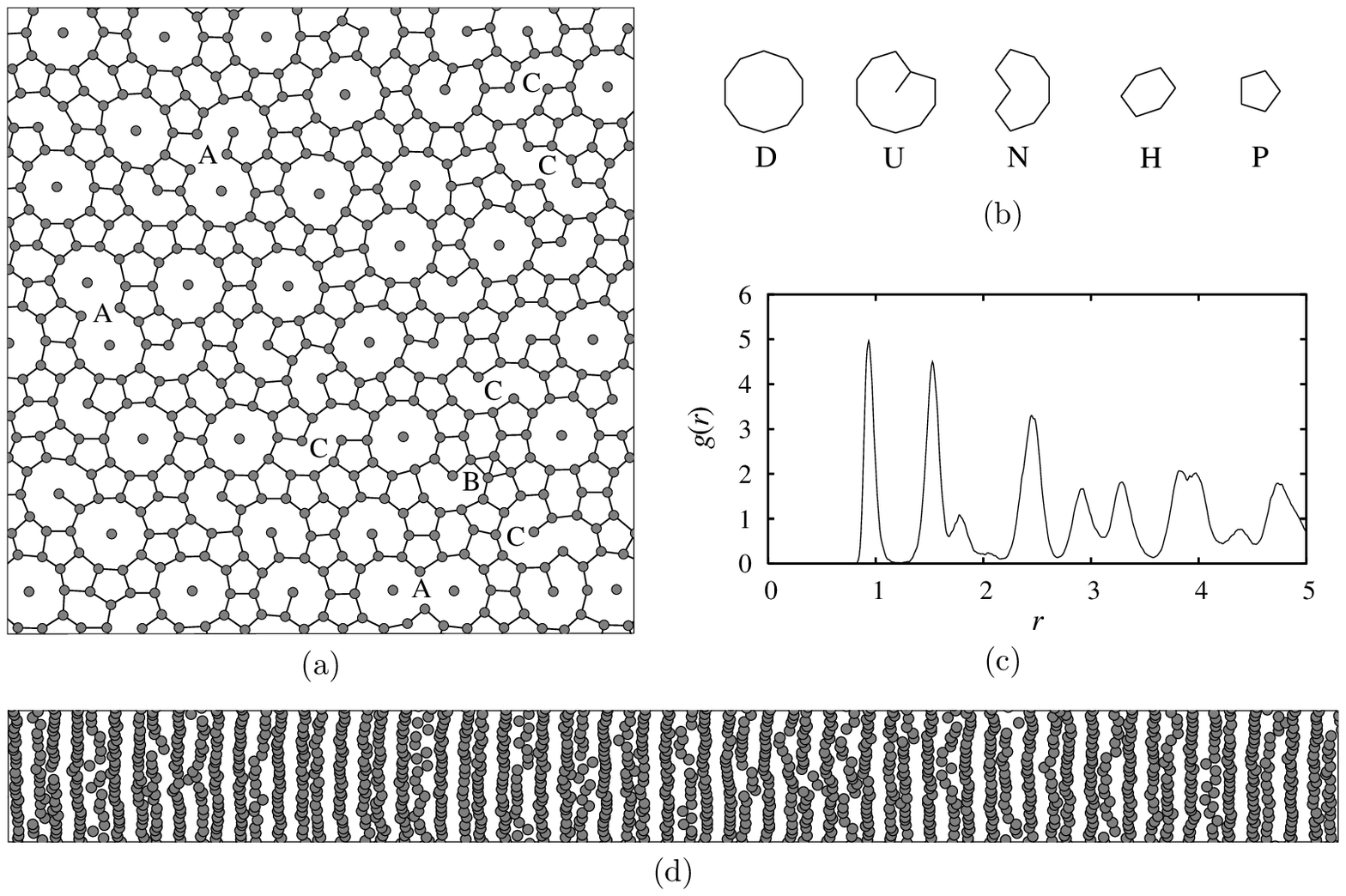}
  \caption{(a)~Snapshot of the decagonal random tiling at $T=0.50$. (b)~The
    five basic tiles are (D)ecagon, (U)-tile, (N)onagon, (H)exagon, and
    (P)entagon. (c)~Radial distribution function $g(r)$. (d)~Oblique view of
    the particle configuration.\label{figure1}}
\end{figure}

The diffraction image of a large sample with $10\;000$ particles, equilibrated
over $5\cdot10^{7}$ molecular dynamics steps, has already been shown in Fig.~4
of Ref.~\cite{Engel2007}. Here, in Fig.~\ref{figure1}(a), we present a
snapshot of the particle configuration. A tiling is drawn by connecting
nearest neighbors. It is found that the majority of the tiles are identical to
one of the five basic tiles shown in Fig.~\ref{figure1}(b). Additionaly,
vacancies (marked as A in Fig.~\ref{figure1}(a)), interstitials (B), as well
as other tiles (C) are frequently found. However, the equilibrium density of
these defects strongly decreases upon lowering the temperature. A small
deformation of the tiles due to phonon motion is also observed. The radial
distribution function in Fig.~\ref{figure1}(c) has peaks near the minima of
the potential at $r_{1}=0.95$ and $r_{2}=1.53$. A third and a fourth smaller
peak are found at $r_{3}=1.79$ and $r_{4}=2.04$. The tiling is algebraically
long-range ordered as seen in the oblique view (Fig.~\ref{figure1}(d)). The
particles align preferably on parallel lines with large ($L$) and small ($S$)
separations, forming a randomized Fibonacci chain. The occurrence ratio of the
separations is $\#L:\#S=40:25=1.6$, i.e.\ close to the golden mean.

For pair interactions, the potential energy per particle is given by the
radial distribution function $g(r)$ and the interaction potential only:
\begin{equation}\label{equation2}
  E\propto\int g(r)V(r)\;r\text{d} r.
\end{equation}
Thus, we can determine $E$ by looking at the peak positions in $g(r)$.  Since
the bonds of the tilings are drawn from the nearest neighbor connections, they
correspond to the first peak at $r_{1}$.  According to (\ref{equation2}), this
peak stabilizes single tiles, but does not prefer any arrangement of the
tiles. It can be shown that the second-, third-, and fourth-nearest neighbor
connections corresponding to $r_{2}$, $r_{3}$, and $r_{4}$ do not cross the
boundaries of the tiles.  Therefore, they also do not link neighboring tiles
energetically.  Further peaks at $r>2.3$ are not important for the
stabilization, because the potential function has already decayed
significantly.  Together, this means that there are no matching rules at all
in the tiling.  The decagonal phase is stabilized completely indirectly by
geometric constraints.

We calculate the potential energy of a hypothetical tiling built from a single
tile only. To do so, we determine the coordination numbers $C_{i}$ for each of
the five basic tiles. $C_{i}$ is equal to the number of neighbors
corresponding to the peak at $r_{i}$. For $C_{1}$, bonds contribute $1/2$ to
avoid counting them twice.  Additionally, $N$ is the effective number of
particles per tile: Each particle on the tile boundary contributes $2\pi/\phi$
to $N$, where $\phi$ denotes the interior angle formed by the tile boundaries
at the position of the particle.  A particle inside the tile contributes a
factor of 1. Using (\ref{equation2}), the potential energy is then given by
\begin{equation}\label{equation3}
  E=\frac{1}{N}\sum_{i=1}^{4}V(r_{i})C_{i}.
\end{equation}

The comparison in Tab.~\ref{table1} shows that the smallest tile $P$ has the
lowest energy and the largest tile $D$ the highest.  However, a periodic
tiling is of course not possible with the $P$-tile alone.  So why is e.g.\ the
tile $D$ formed?  To see this, we take a look at the energy of the decagonal
phase and the approximant Xi, obtained from simulations.  They are $E=-6.66$
and $E=-6.70$, respectively. Both numbers are lower than the values for all
tiles but the $P$-tile. The integration of the $D$-tiles into the tiling seems
to reduce $E$. Paradoxically, as we show later, the tiling with the highest
density of $D$-tiles is the one with the lowest potential energy.  It will
turn out that this is the approximant Xi.
\begin{table}
  \centering
  \tabcolsep5.0mm
  \begin{tabular}{ccccccccc}\hline\hline
    Tile & \multicolumn{4}{c}{Coordination numbers} & Eff. particle & 
    Potential\\
    type & $C_{1}$ & $C_{2}$ & $C_{3}$ & $C_{4}$ & number $N$ &
    energy $E$ \\\hline
    P & 2.5 &  5 &  0 & 0 & 1.5 & -7.85 \\
    H & 3   &  5 &  2 & 2 & 2   & -6.44 \\
    N & 4.5 &  8 &  4 & 2 & 3.5 & -5.90 \\
    U & 6   & 11 &  7 & 2 & 5   & -5.75 \\
    D & 5   & 10 & 10 & 0 & 5   & -5.41 \\\hline\hline
  \end{tabular}
  \\[0.3cm]
  \caption{Comparison of the five basic tiles. $E$ is the potential energy of
    a hypothetical tiling built from a single tile only.\label{table1}}
\end{table}

\section{Potential energy of orthorhombic approximants}

Each approximant of the decagonal quasicrystal corresponds to an average
phason strain $\chi\neq 0$. To show that the phase Xi is the energy ground
state, the potential energy as a function of the phason strains is determined.
We follow \cite{Koschella2002} and generate a large number of approximants
with different values of $\chi$. In the linear theory of elasticity extended
to quasicrystals it can be shown that the decagonal phase has only two
independent phason strains $\chi_{1}$ and $\chi_{2}$~\cite{Bak1985}. Since for
every value of $\chi_{1}$, $\chi_{2}$ an orthorhombic approximant with phason
strains arbitrarily close can be found, we use $\chi_{1}$ and $\chi_{2}$ to
parameterize the approximants. In total 2601 orthorhombic approximants with
500 to 40000 particles are generated for a sufficiently fine mesh in the
$\chi_{1}$-$\chi_{2}$-plane. Each of them is simulated with molecular dynamics
and Monte Carlo alternately, while slowly reducing the temperature from
$T=0.3$ down to $T=0$. This allows to bring the approximant into an energy
minimum within the restriction of the imposed external phason strain.

The use of periodic boundary conditions is necessary to prevent a rotation of
the de Bruijn surface. By identifying opposing sides of the system, the de
Bruijn surface is topologically equivalent to a torus. This means that $\chi$
can only change via topological defects, e.g.\ dislocations, which are found
to not play a role in our simulations.

\begin{figure}[t]
  \centering
  \includegraphics[angle=-90,width=0.9\textwidth]{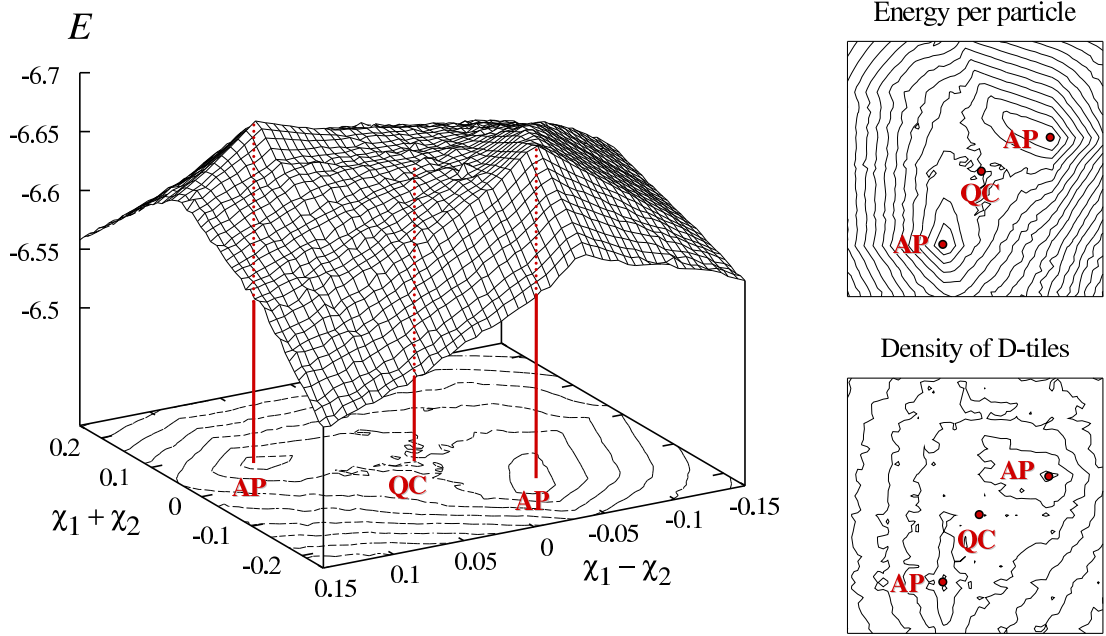}
  \caption{Potential energy $E(\chi_{1},\chi_{2})$ for 2601 orthorhombic
    approximants. Two variants of the approximant Xi have lowest energy,
    forming two peaks in the energy surface (AP). Note the reversal of the
    energy axis. The quasicrystal sits on the saddle point in-between (QC).
    This is also seen in the contour plots shown on the right-hand side.
    \label{figure2}}
\end{figure}

At the end of each simulation, the potential energy is determined. The plot of
$E(\chi_{1},\chi_{2})$ in Fig.~\ref{figure2} confirms that the approximant Xi
is indeed the lowest energy state. Xi appears in two variants: a parallel
arrangement of the supertile rhombs connecting the D-tiles (see
\cite{Engel2007}) at $\chi_{1}=-0.09$, $\chi_{2}=-0.05$ and an alternating
arrangement at $\chi_{1}=0.08$, $\chi_{2}=-0.02$. Note that we do not
distinguish between the variants; together they are called `Xi'. Because the
flipping of a row of supertile rhombs does not cost energy, every tiling built
from the rhombs has the same energy. This is a consequence of the ideal random
tiling character of the decagonal quasicrystal: a rearrangement of the tiling
does not cost energy, if it does not change the tile types.

The random tiling corresponds to the saddle point at $\chi_{1}=\chi_{2}=0$.
Its phason elastic constants are obtained by a quadratic fit. One is positive
and the other one negative: $\lambda_{1}=1.1$, $\lambda_{2}=-2.1$.  High
accuracy is possible by using an adapted selection of approximants with small
phason strains, which we have not done. For stability, both phason elastic
constants have to be greater than zero. The negativity of $\lambda_{2}$
demonstrates once more that the quasicrystal cannot be stable at $T=0$.
Fig.~\ref{figure2} shows that the density of $D$-tiles and the potential
energy are correlated: An increase of the number of $D$-tiles lowers the
energy as shown in the contour plots. This makes sense, since the approximant
Xi is the densest possible packing of the $D$-tiles.

We finish with two remarks: (i)~For large phason strains the decagonal
symmetry is lost and there are four independent phason strains (the most
general case), which means that non-orthorhombic approximants will play a
role. However, since it is not possible to increase the density of $D$-tiles
further, the approximant Xi is also the ground state including
non-orthorhombic approximants. Of course, there are monoclinic variants of Xi.
(ii)~The potential energy in Fig.~\ref{figure2} is not everywhere a smooth
function of the phason strains. Around the origin, $E(\chi)$ is analytic
because the quasiperiodic tiling can compensate an imposed external phason
strain in a continuous manner. In contrast, the two extrema identified as the
approximant Xi are cusps. For larger phason strains there are regions where
$E(\chi)$ is linear. Here, matching rules seem to play a dominant role.
Furthermore, sharp bends are observed in-between.  Simulations of larger
systems are necessary to understand the phason strain relaxation in detail.

Finally, let us come back to the questions raised in the abstract. We have
demonstrated, that the decagonal quasicrystal is an ideal random tiling
without matching rules. It is stabilized by the entropy of phason
fluctuations. At temperatures below $T_{C}$, the approximant Xi is favored,
because it has the lowest energy of all approximants. The transformation
mechanism is a collective rearrangement of the tiling by discrete flips. Our
results confirm that although a locking transition is not possible in two
dimensions, a transition to an approximant can occur very well.

\section{Acknowledgments}

Financial support from the Deutsche Forschungsgemeinschaft under contract
number TR 154/24-1 is gratefully acknowledged.

\end{document}